# Challenges to Keeping the Computer Industry Centered in the US
## DRAFT

*Thomas M. Conte (Georgia Institute of Technology), Erik P. Debenedictis (Sandia National Laboratory), R. Stanley Williams (HP Labs), Mark D. Hill (University of Wisconsin- Madison)*

It is undeniable that the worldwide computer industry's center is the US, specifically in Silicon Valley. Much of the reason for the success of Silicon Valley had to do with *Moore's Law*: the observation by Intel co-founder Gordon Moore that the number of transistors on a microchip doubled at a rate of approximately every two years. According to the *International Technology Roadmap for Semiconductors*, **Moore's Law will end in 2021.** *How can we rethink computing technology to restart the historic explosive performance growth?* Since 2012, the IEEE Rebooting Computing Initiative (IEEE RCI) has been working with industry and the US government to find new computing approaches to answer this question. In parallel, the CCC has held a number of workshops addressing similar questions. This whitepaper summarizes some of the IEEE RCI and CCC findings. **The challenge for the US is to lead this new era of computing.** Our international competitors are not sitting still: China has invested significantly in a variety of approaches such as neuromorphic computing, chip fabrication facilities, computer architecture, and high-performance simulation and data analytics computing, for example. **We must act now,** *otherwise, the center of the computer industry will move from Silicon Valley and likely move off shore entirely.*

Moore's Law produced a golden era where software could be developed independently of computer hardware. Development of software is expensive and requires a highly educated workforce. Being able to run software developed for today's computers on tomorrow's has reigned in the cost of software development. But this guarantee between computer designers and software designers has become a barrier to a new computing era. The "good news" is that there are several ways to restart the meteoric rise of computer performance. But the "bad news" is that the more revolutionary of these approaches will require not only significant hardware investment, but also significant software rewriting. Such change is risky, and unfortunately, *both US software and hardware industries today are risk averse.*

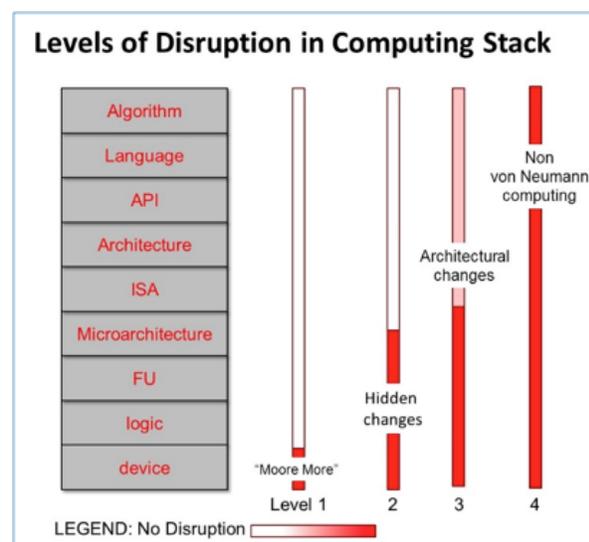

The figure to the right shows the computing stack, from a computer program's algorithm, to its language choice, down through the architecture and ultimately the devices that act as "switches" for the computer's hardware. All approaches to "rebooting computing" can be grouped into four levels based on this figure, where each level provides opportunities but introduces significant challenges. Let's consider each of these levels in turn.

**Level 1: "More Moore:"** This approach is to extend Moore's law past the year 2021 somehow. Although this disrupts the least amount of the computing stack, the challenges are great and potentially insurmountable. The basis for computing since the early 1980s is the CMOS (complementary metal oxide semiconductor) switch. Prior to CMOS, there were other semiconductor switch designs, but CMOS has proven to be the most energy efficient. But we've pushed the envelope for CMOS and its related technologies to its breaking point. The potential for a new, more energy efficient and faster kind of switch is ultimately limited by the laws of physics. Consequently, the semiconductor industry in the US will be able to provide more transistor switches over time, but not significantly faster or better switches than we have today.

**Level 2: Hidden changes:** There is a potential for using novel ways to construct computer internals while only disrupting part of the computing stack. These techniques include adiabatic, reversible, and cryogenic/superconducting computing. Adiabatic and reversible computing exploit the phenomenon in which power in a computer circuit is consumed when the number of inputs is reduced to a smaller number of outputs. Recycling unused inputs can save significant power, but requires very different devices than CMOS. Another approach is superconducting: if certain materials are cooled to very low temperatures (e.g., -452°F), electrons can travel with zero resistance. As with the prior approach, constructing a superconducting computer requires different devices than the semiconductor industry produces today. If either of these examples can be moved from the lab into practice there is a potential for new computers that still run today's software base.

**Level 3: Architectural changes:** A third approach is to allow change to the architecture of the computer. An example is *Approximate* and *Stochastic* computing: computers today oftentimes calculate results to a higher than required accuracy and precision. Removing this waste can both save significant power and improve computing speed. Another approach is to move the program to the data instead of the other way around: today, data is moved in and out of the CPU for computation, but with massive data it makes more sense to move the computation to the data than the other way around. A third approach is to build very specialized computer components that solve a particular problem (i.e., are non-programmable), but do so at much higher efficiency. None of these approaches will speed up today's software. They will all require significant investment in both hardware *and* new software. But the potential for a new era of expanded computing performance is much higher than provided by level 2.

**Level 4: Non-von Neumann:** The current way we compute was first articulated by John von Neumann in 1948. But there are radically different ways to compute that may be significantly better. For example, *Quantum Computing* uses properties of quantum mechanics to solve problems far more quickly than the von Neumann approach. Quantum computing is not a universal computing platform—it solves a limited set of problems, but the problems are very important to science, engineering and national security. For example, a quantum computer would be able to factor the product of two large primes in a nanosecond. This is significant because *asymmetric key encryption* is based on the principle that such factoring is intractable computationally. Every facet of e-commerce and national security relies on this encryption standard. Another level-4 approach is to build *Neuromorphic Computers* that leverage what is known about how the human brain operates. For example, a neuromorphic computer is well suited for recognizing and classifying patterns in text, audio or images. It's no surprise when you think of it: the human brain is remarkably efficient at such tasks. Both quantum and neuromorphic computers require significant investment in all levels of the computing stack. However, the performance potential is much higher than any level 1-3 approaches.

*In order for the US to continue its historic leadership role in the computer industry, it will require significant investment in research, development and manufacturing to bring the most promising approaches forward.* The National Strategic Computing Initiative (NSCI) executive order (signed by President Obama on July 29, 2015) is a solid first step, but it needs to be expanded beyond its focus on high-performance computing. Meanwhile, other nations such as China and the EU are not sitting still. Our risk-averse US computer industry will continue to invest in level-1 approaches only. Without significant investment of our own, the center of the computing industry will likely shift from the US to elsewhere. We must make significant injection of new funding for approaches across levels two, three and four. NSCI identified five agencies: DOE, DARPA, NSF, IARPA and NIST. Critically, none of the missions of these agencies are aligned with the goals of keeping the computing industry in the US. A second component for success must include creation of a central coordinating office with the goal of keeping the US in the lead in the next revolution in computing.

*This material is based upon work supported by the National Science Foundation under Grant No. 1136993. Any opinions, findings, and conclusions or recommendations expressed in this material are those of the authors and do not necessarily reflect the views of the National Science Foundation.*